\documentclass[10pt,tighten,aas2pp4]{aastex}

\begin{document}
\title{Imaging extrasolar planets by stellar halo suppression in separately-corrected
color bands}
\author{Johanan L. Codona and Roger Angel}
\affil{\textit{Steward Observatory, University of Arizona}}

\begin{abstract}
Extra-solar planets have not been imaged directly with existing ground
or space telescopes because they are too faint to be seen against
the halo of the nearby bright star. Most techniques being explored
to suppress the halo are achromatic, with separate correction of diffraction
and wavefront errors. Residual speckle structure may be subtracted
by differencing images taken through narrowband filters, but photon
noise remains and ultimately limits sensitivity. Here we describe
two ways to take advantage of narrow bands to reduce speckle photon
flux and to obtain better control of systematic errors. Multiple images
are formed in separate color bands of 5-10\% bandwidth, and 
recorded by coronagraphic interferometers equipped with active
control of wavefront phase and/or amplitude. In one method, a single
deformable pupil mirror is used to actively correct both diffraction
and wavefront components of the halo. This yields good diffraction
suppression for complex pupil obscuration, with high throughput over
half the focal plane. In a second method, the coronagraphic interferometer
is used as a second stage after conventional apodization. The
halo from uncontrollable residual errors in the pupil mask or
wavefront is removed by destructive interference
made directly at the detector focal plane with an ``anti-halo'',
synthesized by spatial light modulators in the reference arm of the
interferometer. In this way very deep suppression may be achieved
by control elements with greatly relaxed, and thus achievable, tolerances.
In both examples, systematic errors are minimized because the planet
imaging cameras themselves also provide the error sensing data.
\end{abstract}


\section{INTRODUCTION}

Optical imaging of extrasolar planets, whether from space or
the ground, requires improved methods for halo suppression. These
need to work over a broad wavelength band to minimize photon noise
in both the wavefront sensor and planet images. Currently the best controlled
starlight halos are obtained at $\sim$$1.65$ $\mu$m wavelength from both 
space and ground, with speckle structure at $\sim$$2.5\times10^{-4}$
of the star intensity at $\sim$$0.5$ arcsec radius (Schneider,
2002, Close et al, 2003). Subtraction methods such as rotation of
the Hubble Space Telescope (HST) and spectral differencing for ground
telescopes (Racine et al, 1999) yield a detection limit for faint
companions at $\sim$$10^{-5}$ contrast. Order-of-magnitude improvements
of these methods may be adequate to image massive, young giant planets
in wide orbits, but older and less massive giants will generally be much fainter, 
even when heated by the star. Fainter still will be terrestrial planets, thus 
an Earth in a twin of the solar system at 10 pc would appear
in reflected light at $\sim$$ 2\times10^{-10}$ of the sun at 0.1 arcsec maximum
separation.  Detection at such faint
contrast will likely require new and more powerful techniques, to
directly reduce the photon flux (and noise) of the speckle halo, while
maintaining exquisite control of systematic errors. 

Most proposed optical configurations for deep halo suppression are
inherently achromatic, involving a combination of very efficient apodization
or coronagraphy (Kasdin et al, 2003, Guyon 2003 and Gonsalves and
Nisenson, 2003) with active correction of wavefront errors (Trauger
et al., 2003), and, in some concepts, also of pupil amplitude (Littman
et al., 2003). For imaging very close to the star, where terrestrial
and old giant planets are likely to be found, the halo component from
optical errors is set by lower-order terms in the wavefront aberration,
and these can be corrected by deformable mirrors of high accuracy
but modest spatial resolution (Malbet, Yu \& Shao, 1995). However,
if wavefront measurements are made across the pupil to determine even
low-resolution correction coefficients, they would need to be both
finely detailed and highly accurate, because any under-sampled high-order
structure will alias to low order. 

This difficulty can be overcome if wavefront and amplitude errors
are derived from focal plane measurements. Thus, improved measurement
accuracy for phase and amplitude corrections to be applied at the
pupil is obtained by phase diversity techniques applied to focal plane
images (Malbet et al, 1995, Jefferies et al. 2002). This method was
used to measure in detail the HST primary mirror aberrations (Roddier
\& Roddier, 1993). All focal plane techniques require the weak
halo fluxes to be sensed in limited wavelength bands, and multiple channels will 
be needed to improve the sensitivity set by photon noise. When rapid
corrections are based on wavefront measurements made in the pupil plane,
the limit to wavefront accuracy set by photon noise results in a halo with very
weak speckles containing just a few photons each (Angel, 1994). It
can be shown that focal plane measurements of the weak halo speckles
themselves can give the same accuracy and level of suppression, provided
the same total bandwidth is used (Angel, 2003). Focal plane measurements
have the advantages that accurate calibration is not needed, because
the measurement is null, and non-common-path errors are eliminated
if the same detectors are used to provide the deep integrated planet
images as well as instantaneous wavefront data. For our application
in which the star image must be blocked, it is convenient to derive
the errors interferometrically rather than by phase diversity, obtaining
the complex amplitude of the weak residual halo with the aid of a
reference beam derived from the blocked starlight (Angel, 2003).


\begin{figure}
\figurenum{1}
\epsscale{0.9}
\plotone{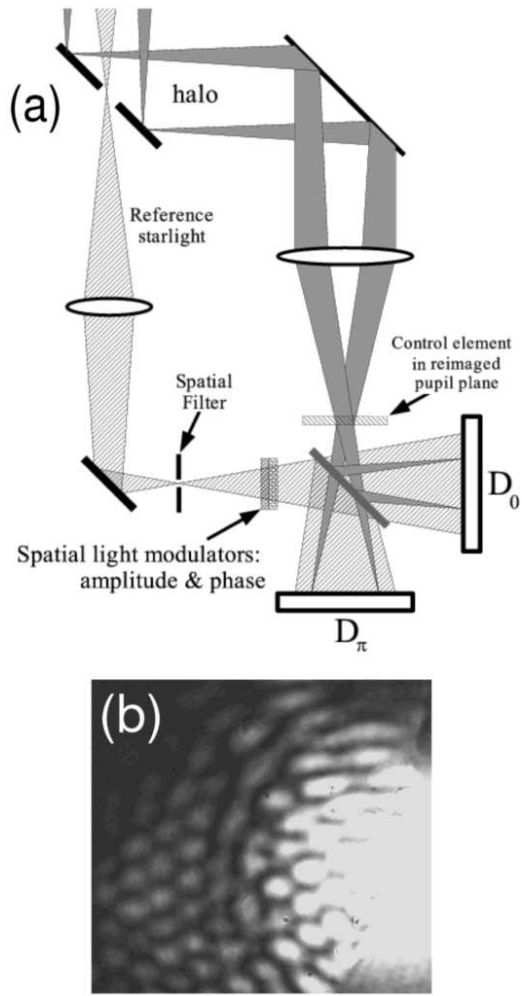}
\caption{Concept for a narrow-band coronagraphic
interferometer to sense and suppress the stellar halo. (a) A narrowband 
stellar image enters the system from above. The
halo and planet image are reflected from a pierced mirror to be relayed via a pupil
stop and beamsplitter onto the photon-counting detector arrays marked
$\mathrm{{D}_{0}}$ and $\mathrm{{D}_{\pi}}$. The diffraction-limited
core is passed through the mirror and a spatial filter to be used
for phase reference and/or destructive interference of the halo. Active
or passive phase and amplitude control elements (e.g. a deformable
mirror or spatial light modulator) may be placed in the halo light
path at the reimaged pupil (Lyot stop), or in the reference beam.  
(b) part of Airy 
pattern recorded in a lab demonstration, with the reference 
wavefront slightly tilted so that
constant phase in the halo would appear as horizontal fringes. The
fringe displacement of successive Airy rings reveals the half wave phase
shift between them.}
\end{figure}

\section{HALO SUPPRESSION IN NARROW BANDS}

\newcommand{\Deff}{D_{\mathrm{eff}}} 

Since multiple wavelength channels are desirable for accurate wavefront
sensing, it makes sense to explore the potential for passive and active
corrections within each limited bandwidth channel to directly suppress
the halo. New modes then become possible, because separate correction
of diffraction effects and wavefront errors is no longer required.
To implement such schemes, an initial separation of the telescope
image into multiple wavelength star images would be made with cascaded
dichroic beamsplitters or a field grating (Bonnet \& Court\`{e}s 1962).
Labeyrie (2002) has pointed out one approach, in which a conventional
coronagraph is followed by a second stage reimaging device. In it
the halo speckle phases are actively modified to be all the same,
so at the center of an intermediate quasi-pupil their energy is concentrated
in a central spot that can be blocked. In our concept for suppression,
a coronagraphic interferometer is used to both measure and suppress
the halo in a servo loop, as shown schematically in figure 1. The
active elements to control phase and/or amplitude are placed at the
pupil image formed in the signal arm, or in the reference beam arm
to modulate the reference image. One such instrument would be placed
at each narrowband star image. In operation at the telescope, the
corrections to be applied are determined from measurements of the
complex amplitude at each focal plane image pixel, obtained from images
recorded before and after a phase shifts of $\pi/2$ in uniform amplitude
reference beams.

Multiple instruments, each with two-dimensional arrays of active elements,
may at first seem a daunting proposition, but suitable small MEMS
deformable mirrors and liquid-crystal spatial light modulators for
intensity and phase control (Littman, 2003) are becoming available,
as are CCDs with gain for use as fast, photon-counting arrays (Mackay,
2001). We consider below two new modes of operation. The first is
a phase-only alternative to separate apodization and wavefront correction.
In the second, conventional apodization and wavefront correction is 
followed by a second stage in which the residual halo is further reduced 
by destructive interference.

\section{METHOD I: PHASE CORRECTION OF BOTH DIFFRACTION HALO AND WAVEFRONT
ERRORS}

In this method, the full halo (including diffraction rings) is suppressed
not by apodization or Lyot stop, but by modulation of phase across
the pupil. The basic idea is to superpose Fourier phase components
across the pupil that create the complex amplitude at each spot in
the focal plane needed to null out the measured value. This is not
possible over the whole focal plane, as can be seen from symmetry
considerations. Thus the complex amplitude of the diffraction halo
is Hermitian, while from pupil phase modulation it is anti-Hermitian,
so correction can only be made over half the focal plane. Also, the
suppression is more efficient if the outer radius is limited, so we
adopt a {}``D''-shaped nulled region in the focal plane. The required
pupil phase shifts are found by iterative Fourier transform of the
complex halo amplitudes measured within the D, and are of order 1
radian. We envisage that the phase pattern needed to suppress diffraction
would be figured permanently on the surface of a deformable mirror,
and the active control used to clean up residual errors. Phase shifts
made by displacement 
to correct diffraction 
can be accurate only at one wavelength $\lambda_{0}$,
and will be systematically off at neighbouring wavelengths. 
For reasonably
small changes in wavelength, the relative phase error is given by
$\Delta\phi/\phi = (\lambda-\lambda_{0})/\lambda_{0}$,
therefore there will be a chromatic increase in the halo intensity
that varies as  
$((\lambda-\lambda_{0})   /\lambda_{0})^{2}$ times the original halo intensity.


\begin{figure}[t]
\figurenum{2}
\epsscale{1.0}
\plotone{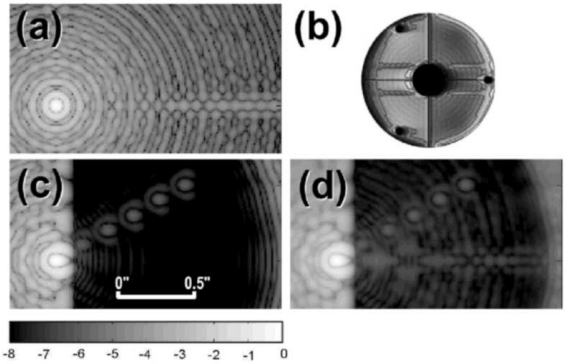}
\caption{Correction of the HST diffraction
halo by pupil phase adjustment.  (a) The diffraction pattern at 
wavelength 0.5$\mu$m over 8 decades, given ideal
figure correction. (b) Pupil phase change required to correct for diffraction, optimized
for field radius 2.5 -- 35 $\lambda/D$. The contour interval is $\pi$/4. (c) Monochromatic 
PSF after correction. Companions at $10^{-6}$ star are revealed at radial separations
in 0.2 arcsec.  (d) PSF at spectral bandwidth of 10\%. }
\end{figure}

As an illustration, we show this method applied to the HST in figure
2. Obscuration of the primary mirror by the secondary and supports
results in a complex diffraction pattern. The optimized pupil phase
modification yields about four orders of magnitude reduction for monochromatic
light, with only modest loss of energy (the Strehl ratio is 43\%).
We have modeled the reduced halo suppression for finite bandwidth
of 10\%. The contrast at
radius $3.5\lambda/D$ (0.15 arcsec at $\lambda=0.5\mu m$) averaged over the band 
is $3\times10^{-7}$,
still 3000 times below the uncorrected diffraction level.

Such deep suppression and good throughput close to the star compares
favorably with conventional coronagraphy. HST's NICMOS coronagraph
with a hard field stop of radius 2.1 $\lambda/D$ allows detection
of companions as close as 2.6 -- 3.7 $\lambda/D$, but with suppression
of only $2.5\times10^{-4}$ (Schneider 2002). A more aggressive coronagraph
for HST, including optimized correction of low-order optical errors,
could 
yield strong suppression of $3\times10^{-8}$, but at larger radius
(7.5 $\lambda/D$) and with only 18\% planet transmission (Malbet
et al. 1994, 1995). A valuable aspect of the new method is that its
deep suppression of diffraction and scattered halo can be maintained
by adjustments with deformable mirrors alone, avoiding the reliance
of conventional coronagraphs on precise manufacture and placement
of a pupil mask. It will be particularly powerful for complex or obstructed
pupils for which apodization and Lyot stop methods are not very efficient
(Sabatke et al, 2003). For example, it could be used to correct the
halos caused by gaps and phase steps at the segment boundaries of
the James Webb and Keck telescopes.

\section{METHOD II: ACTIVE HALO NULLING}

For deepest suppression over the full annular aperture, apodization
would be combined with a second operational mode we now describe.
The idea is to suppress the residual halo directly, by destructive
interference in the focal plane. For this mode, the interferometer
of figure 1 would be configured as a coronagraph, with halo suppression
achieved in the first instance by apodization preceding the interferometer.
Based on a Fourier transform of the star halo's measured complex amplitude,
the deformable mirror at the Lyot stop within the coronagraphic interferometer
would be adjusted to reduce the halo from phase errors. Some faint
speckles will remain, because of limits to the manufacturing accuracy
of the field and Lyot stops and in the setting of the phase correction
device. Their complex amplitudes would be re-measured, 
and the reference beam would be modulated 
to create a corresponding ``anti-halo'' 
with the same amplitudes but opposite phases.  
Destructive interference
would then further reduce the halo. A planet image would be interpreted
by the system as a speckle, but could not be suppressed, since it
is incoherent with the starlight. 
Techniques for complex amplitude control with deformable mirrors
and spatial light modulators are described by Gonsalves, 1997, Roggemann \& Lee,
1997 and Littman et al 2003. 

The additional reduction achievable in this way depends on bandwidth
because of the radial stretching of speckles with wavelength. Remembering
that phase changes by $\pi$ for a radius change of $\lambda/\Deff$
(figure 1b), we find $I/I_{0}\approx(\pi\phi/2)^{2}(\Delta\lambda/\lambda_{0})^{2}$,
where $\Delta\lambda$ is the full bandwidth and $\phi$ is the field radius
in units of $\lambda/\Deff$. $\Deff$ is the effective aperture for
resolution after apodization.

The value of this second approach is that the accuracy now required
for additional wavefront and amplitude control is much relaxed, because
the complex amplitude to be adjusted is that of the weak interfering
beam, not the full starlight beam. Thus, no matter how faint the starting
halo, if we desire additional suppression by a factor of 100, the
accuracy required in measuring and setting the anti-halo is only $\sim$$0.1$
radians in phase and 10\% in amplitude. Such modest requirements are
readily achievable with liquid crystal modulators of limited phase
resolution.


\begin{figure}[h]
\figurenum{3}
\epsscale{1.0}
\plotone{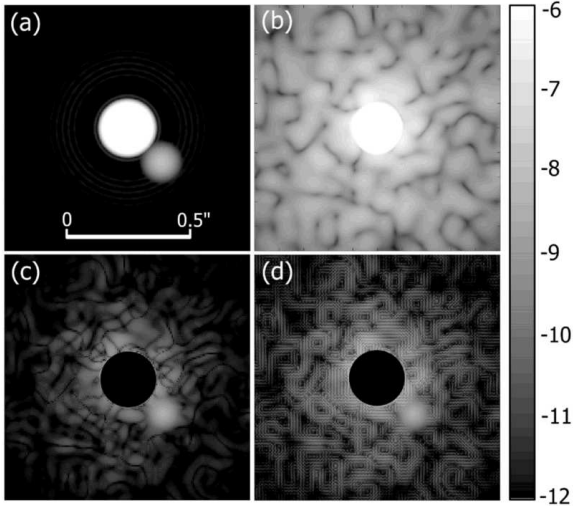}
\caption{Halo suppression
by destructive interference in the focal plane.  (a) Intensity of a
star halo ($\log_{10}$ grayscale running from $10^{-6}$--$10^{-12}$
peak) calculated for an ideal, deep, radial Chebyshev apodization
(IEEE 1979) of an unobstructed circular aperture, adjusted for a halo
at $10^{-12}$ level, with a companion of $10^{-8}$ star and 0.2 arcsec
separation.  This mask has transmission of 7.5\% and doubles the
diffraction-limited image width. The scale is shown for $D=4$ m
($\Deff \sim 2$ m) and $\lambda=500$ nm.  (b) The halo modelled for
predominantly low-order residual errors in the mask (power spectrum as
$\kappa^{-5/3}$ and strength described in text).  (c) The halo now
reduced by destructive interference with an ``anti-halo'', modeled for
5\% bandwidth as described in the text.  (d) As (c), but including the
effect of a spatial light modulator with discrete pixels subtending
$\lambda/3D$.}

\end{figure}
Figure 3 shows an example of suppression at the level required for
detection of close-by Earth-like planets. The case modelled is a 4 m
space telescope, heavily apodized for $<10^{-12}$ suppression, but
degraded by residual amplitude and phase errors that result in a halo
of $2\times10^{-7}$ at 0.2 arcsec. When the focal plane nulling is
applied, cancellation to a level of $3\times10^{-10}$ is obtained,
even allowing for 5\% bandwidth. To minimize photon and speckle noise,
measurements from 6 to 10 bands would be combined. Optimum settings
for the global and local correction elements would be worked out together.
In this example, the required accuracy for the ``anti-halo'' setting
is $> 100$ times more relaxed than that assumed for the initial direct
pupil correction (0.2 nm rms in wavefront and 0.25\% in intensity
at 0.25 m scale).

\section{DISCUSSION}

In general, a servo correction system will be necessary to repeatedly 
correct small distortions that develop in the wavefront. Only a few photons 
per speckle need be recorded over the full 
spectral band before a useful correction can be
made.  Repeated corrections made in this way will ensure that the halo flux
is at the minimum possible level, and since each speckle is corrected
before it has a chance to build up to more than a few photons, speckle noise
in each cycle is held close to the level of halo photon noise.  

This approach is valid for both ground and space telescopes.  The main
differences are in the achievable suppression limit and in the servo
cycle time (Angel, 2003).  For ground telescopes, the wavefront is
continuously perturbed by evolving atmospheric turbulence.  While some
evolution can be predicted and compensated, there will remain an
unpredictable component that causes speckles at the few photon level
to develop on a scale of a few tenths of a millisecond, depending on
telescope aperture and seeing conditions.  The corresponding halo
brightness for telescopes of 8 to 30 m aperture at $\sim$$1$ micron
wavelength is likely to be $\sim$$10^{-6}$--$10^{-7}$ of the star. For
optical coronagraphs in space, the relative stability of the wavefront
will permit much better halo suppression. But once the halo is at a
level below that of the zodiacal background, there is little advantage
in going further.  For a telescope of a few meters aperture imaging at
$\sim$$0.5$ microns, this limit corresponds to $\sim$$10^{-9}$ of the star.  In
the photon noise limit, limiting sensitivity to exoplanets varies as
$($halo intensity$) \times \surd($cycle time$)$, 
and is comparable for these
ground and space telescope examples.

The level of suppression needed for ground telescopes should be achievable
by method I with a single very fast-acting deformable mirror.  The
method has the advantage, mentioned above, that complex diffraction
spikes from very large segmented mirrors can be suppressed with little
loss of light.  For the much higher suppression needed in space, the
active halo nulling method will be preferred.

\acknowledgements{This research is supported by AFOSR contract F49620-01-0383 and the
NSF under grant AST-0138347.}

\newpage

\end{document}